\pdfoutput=1
\documentclass[aps,prc,twocolumn,floatfix,superscriptaddress,nofootinbib,preprintnumbers,longbibliography,amsmath,amsthm,amssymb]{revtex4-2}
\usepackage{graphicx}
\usepackage{amsfonts}
\usepackage{color}
\usepackage[colorlinks=true,linkcolor=blue,citecolor=blue]{hyperref}
\usepackage{dcolumn}
\usepackage{bm}
\usepackage{braket}
\usepackage[normalem]{ulem}  

\begin{document}
\allowdisplaybreaks[1]
\title{Evolution of giant monopole resonance with triaxial deformation}
\author{Kouhei Washiyama}
\email[E-mail: ]{washiyama@nucl.ph.tsukuba.ac.jp }
\affiliation{Center for Computational Sciences, University of Tsukuba, Tsukuba, Ibaraki 305-8577, Japan}
\affiliation{Research Center for Superheavy Elements, Kyushu University, Fukuoka 819-0395, Japan}
\author{Shuichiro Ebata}
\email[E-mail: ]{ebata@mail.saitama-u.ac.jp }
\affiliation{Graduate School of Science and Engineering, Saitama University, Saitama 338-8570, Japan}
\author{Kenichi Yoshida}
\email[E-mail: ]{kyoshida@rcnp.osaka-u.ac.jp}
\affiliation{Research Center for Nuclear Physics, Osaka University, 
Ibaraki, Osaka 567-0047 Japan}
\affiliation{Center for Computational Sciences, University of Tsukuba, Tsukuba, Ibaraki 305-8577, Japan}
\affiliation{RIKEN Nishina Center for Accelerator-Based Science, Wako, Saitama 351-0198, Japan}
%
\date{\today}
\begin{abstract}
\begin{description}
\item[Background]
The isoscalar giant monopole resonance (ISGMR) splits
into two peaks in prolately deformed nuclei.
When a nucleus is triaxially deformed,
a peak appears in the middle between the two peaks.

\item[Purpose]
We investigate the mechanism of the appearance of the middle peak in the ISGMR in triaxial nuclei.

\item[Method] 
We perform the constrained Skyrme-Hartree-Fock-Bogoliubov (CHFB) calculation for arbitrary triaxial shapes in $^{100}$Mo.
We calculate the strength functions of the isoscalar monopole (ISM) and IS quadrupole modes 
on the CHFB states.
Furthermore, 
we investigate vibrations of matter distributions in $x$, $y$, and $z$ directions induced by the external ISM field,
with the $z$ axis being the longest axis of the triaxial shape.

\item[Results]
The middle peak in the ISM strength
evolves from the triaxial degree $\gamma=0^\circ$ to $60^\circ$.
This is because 
the difference between 
the vibration in $x$ direction and that in $y$ direction
is evident
with an increase in $\gamma$ and
the quadrupole $K=2$ component of the induced density of the ISM at the middle peak increases as $\gamma$ increases, where $K$ denotes the $z$ component of the angular momentum.
This property is also obtained in the unperturbed ISM strength without the residual fields. 

\item[Conclusions]
The mixing between the monopole and quadrupole modes is primarily determined by the ground-state deformation. 
Therefore, 
the ISM strength of the middle peak becomes strong 
as the triaxial degree in the ground state increases.

\end{description}
\end{abstract}

\maketitle

\section{Introduction}\label{introduction}
%
Understanding complex behaviors of atomic nuclei lies at 
the heart of low-energy nuclear physics. 
One of the central themes in this domain 
is the investigation of the collective modes 
exhibited
in responses to nuclei.
These collective modes play a pivotal role 
in elucidating the underlying structures 
and dynamics of nuclei, 
thereby providing crucial insights into their fundamental properties, such as nuclear shapes and deformations. 

%
Giant resonances are a fundamental collective mode of excitation in nuclei~\cite{harakeh2001giant}.
It is well known that 
the isoscalar giant monopole resonance (ISGMR) splits into two peaks in prolately deformed nuclei~\cite{GARG201855}.
This was first observed in the well-deformed $^{154}$Sm nucleus~\cite{garg80}.
Prior to this observation, the possible emergence of the two-peak structure in the ISGMR strengths in deformed nuclei had been a subject of discussion~\cite{zawischa78}.
Subsequently, it was found that 
the energy of the lower peak in the isoscalar monopole (ISM) strengths 
coincides with 
that of the $K=0$ component of the isoscalar quadrupole (ISQ) strengths
in prolately deformed nuclei,
where $K$ denotes the $z$ component of the angular momentum.
Recent calculations based on the Hartree-Fock-Bogoliubov (HFB) method plus quasiparticle random-phase approximation (QRPA)
in deformed nuclei
confirmed this coincidence in energy~\cite{Yoshida:2010zz,yoshida10,niksic13,kvasil16,Sun:2022gdu}.

%
Experimental identification of static triaxial deformations 
in nuclei can be achieved through various means, 
such as the observation of chiral doublet bands~\cite{FRAUENDORF1997131} and the wobbling mode, 
which was recently reported in $^{163}$Lu~\cite{PhysRevLett.86.5866}. 
In the low-lying spectra, 
the presence of the $2_2^+$ state 
and the $\gamma$ vibration provides evidence of a certain degree of triaxiality in nuclei.
%
%
As a beyond mean-field method,
the generator-coordinate method (GCM)~\cite{bender08,rodriguez10,yao10,kimura13,suzuki21} and
the collective Hamiltonian method~\cite{libert99,prochniak04,niksic09,Matsuyanagi:2016gyp,washiyama23}
have included the triaxial degree of freedom
as a collective coordinate
and produced a better reproduction of 
experimental low-lying spectra.
Besides that, it has been proposed that high-energy heavy-ion collisions 
can be used to extract information on triaxiality~\cite{Jia:2021qyu}.

%
Recently, Shi and Stevenson \cite{Shi_2023} investigated the ISGMR in $^{100}$Mo with the time-dependent density functional theory (TDDFT) + Bardeen-Cooper-Schrieffer (BCS) method with the Skyrme EDFs.
Then, they found that 
the calculation incorporating a triaxial shape ($\beta\approx 0.28$ and $\gamma=20^\circ\textrm{--}30^\circ$) 
with the SkM$^*$ EDF 
opportunely
reproduced the experimentally observed ISGMR at RCNP~\cite{HOWARD2020135608}.
Notably, the inclusion of finite triaxiality resulted in 
an increase in the ISM strength within the energy range of $E=14$--15 MeV, 
making an additional peak between the two existing peaks and
providing a closer match to the experimental data within that energy range. 
This observation serves as evidence of a static triaxial deformation ($\gamma=20^\circ\textrm{--}30^\circ$) in $^{100}$Mo. 
The authors further investigated variations in the isoscalar quadrupole moments 
over time subsequent to the ISM boost.
%
They conjectured that the appearance of the middle peak is due to a coupling between the ISM and ISQ $K=2$ modes and anticipated that, at triaxial shapes in the ground state, 
the external ISM field induces different vibrations in the three principal axes.

%
The operator of the ISM, $r^2=x^2+y^2+z^2$, is isotropic.
Nevertheless, the reason why the ISM external field 
induces anisotropic oscillations in deformed nuclei remains unclear. 
This study aims to explore the mechanism behind the emergence of the 
middle
peak in the ISM strength 
and to elucidate the origin of anisotropic oscillations triggered by the isotropic ISM field in triaxially deformed nuclei.
We perform 
the HFB + QRPA calculation 
to obtain the response function of 
the ISM 
for arbitrary triaxial shapes in $^{100}$Mo.
For the QRPA calculations, the finite amplitude method (FAM)~\cite{nakatsukasa07} is used
to provide the response function in a numerically feasible way.
Then, we analyze the vibrations of matter distributions in the $x$, $y$, and $z$ directions 
to delve into the underlying mechanisms contributing to the anisotropic oscillations induced by the ISM field.

The article is organized as follows.
Section~\ref{method} describes the method of HFB + FAM-QRPA.
In Sec.~\ref{result},
results of the ISM strengths in triaxial shapes and the analysis 
are presented.
Finally, Sec.~\ref{summary}
summarizes the present article.

\section{Method}\label{method}

We briefly recapitulate the formulation
of the constrained HFB (CHFB) + FAM-QRPA approach, 
and the details can be found in 
Refs.~\cite{nakatsukasa07,stoitsov11,niksic13,washiyama17,washiyama21}.

The FAM equation in the quasiparticle basis is given as 
\begin{subequations}\label{eq:FAM_XY}
\begin{align}
(E_\mu + E_\nu -\omega)X_{\mu\nu}(\omega) + \delta H^{20}_{\mu\nu}(\omega) &= -F^{20}_{\mu\nu} \; ,\\  
(E_\mu + E_\nu +\omega)Y_{\mu\nu}(\omega) + \delta H^{02}_{\mu\nu}(\omega) &= -F^{02}_{\mu\nu}\; ,
\end{align}\end{subequations}
where $E_{\mu}$ are the quasiparticle energies,
$X$ and $Y$ are the FAM amplitudes at a given frequency $\omega$, and 
$\delta H^{20(02)}$ and $F^{20(02)}$ are 
the two-quasiparticle components of an induced Hamiltonian and an external field $\hat{F}$, respectively.
The FAM equation \eqref{eq:FAM_XY} can be solved iteratively at each $\omega$ 
by replacing $\omega$ with complex frequencies $\omega \to \omega + i\Gamma/2$,
where the imaginary part $\Gamma$ corresponds to a smearing width.

From the converged $X_{\mu\nu}(\omega;\hat{F})$ and $Y_{\mu\nu}(\omega;\hat{F})$ amplitudes induced by the external field $\hat{F}$,
the FAM response function, 
the change of quantity associated with
another field $\hat{F}^\prime$ in the response to the perturbation $\hat{F}$, 
is defined as 
\begin{align}
  R_{\hat{F}^\prime\hat{F}}(\omega) &= \sum_{\mu < \nu} [F_{\mu\nu}^{\prime 20*} X_{\mu\nu}(\omega;\hat{F})+ F_{\mu\nu}^{\prime 02*} Y_{\mu\nu}(\omega;\hat{F})].
  \label{eq:FAM_response_function}
\end{align}
Setting $\hat{F}=\hat{F}^\prime$,
the strength function is given as 
\begin{align}
  S(\hat{F};\omega) &= -\frac{1}{\pi}\text{Im} R_{\hat{F}\hat{F}}(\omega).
  \label{eq:FAM_strength_function}
\end{align}

As an external field operator $\hat{F}$,
we employ the one-body isoscalar monopole and quadrupole operators expressed as
\begin{equation}\label{eq:multipole}
    f_{00} = \sum_{i=1}^{A}  r^2_i, \quad f_{2K} = \sum_{i=1}^A r_i^2 Y_{2K}(\hat{r}_i).
\end{equation}
We define the quadrupole operators with the $x$-signature quantum number of $r_x=\pm 1$ 
as $Q^{(+)}_{20} = f_{20}$ and $Q^{(\pm)}_{2K}=(f_{2K}\pm f_{2-K})/\sqrt{2}$ for $K>0$.
These quadrupole operators can also be expressed in terms of the Cartesian coordinate as
\begin{align}
    Q^{(+)}_{20} &= \sum_{i=1}^{A}\sqrt{\frac{5}{16\pi}} (2z_i^2-x_i^2-y_i^2), \\
    Q^{(+)}_{22} &= \sum_{i=1}^{A}\sqrt{\frac{15}{16\pi}}(x_i^2-y_i^2).
\end{align}

We define the quadrupole deformation parameters $\beta$ and $\gamma$ as 
\begin{align}
    \beta& =\sqrt{\frac{5}{16\pi}} \frac{4\pi}{3R^2A} \sqrt{\braket{Q^{(+)}_{20}}^2+\braket{Q^{(+)}_{22}}^2},\\
\gamma &=\arctan\left(\frac{\braket{Q^{(+)}_{22}}}{\braket{Q^{(+)}_{20}}}\right),
\end{align}
where $R=1.2A^{1/3}$\,fm and $A$ is the mass number.
To investigate the evolution of the giant monopole resonance with triaxial deformation,
we vary the triaxiality $\gamma$ from 
$\gamma=0^\circ$ to $60^\circ$, 
where $\gamma=0^\circ$ ($\gamma=60^\circ$) corresponds to the axially symmetric prolate (oblate) shape with the $z$ axis ($y$ axis) being a symmetry axis.
In the region of $0^\circ\le \gamma \le 60^\circ$,
the $z$ and $y$ axes are the longest and shortest axes, respectively.

To prepare triaxial CHFB states,
we solve the Skyrme CHFB equations with the two-basis method \cite{gall94,terasaki95}
in a three-dimensional Cartesian mesh.
The single-particle wave functions in the Hartree-Fock (HF) basis 
are chosen as eigenstates of parity, $z$ signature, and $y$-time simplex
\cite{bonche85,dobaczewski00,ev8new}. 
With this choice, nuclear shapes are obtained with
$x=0$, $y=0$, and $z=0$ plane symmetries and 
the model space is reduced to 1/8 of the full box.
We use a (13.2\,fm)$^3$ box in $x>0$, $y>0$, and $z>0$
with a mesh spacing of 0.8\,fm.
The single-particle basis consists of 
1400 neutron and 1120 proton HF-basis states,
which approximately correspond to the maximum quasiparticle energy of $60$\,MeV.
The constraint quantities are the isoscalar quadrupole moments with $K=0$ and $K=2$, $Q_{20}^{(+)}$ and $Q_{22}^{(+)}$.
We employ the SkM$^*$ EDF \cite{bartel82} and 
the contact volume-type pairing with a pairing 
window of 20\,MeV above and below the Fermi energy in the HF basis as described in Refs. \cite{bonche85, rigollet99,ev8new}.
The pairing strengths are adjusted to 
reproduce the empirical neutron and proton gaps in $^{106}$Pd, 
and the resultant values are 
240\,MeV\,fm$^3$ and 285\,MeV\,fm$^3$ for neutrons and protons, respectively.
We confirmed that change of the pairing strength in 5\% does not affect the conclusions of this article.

We perform the FAM-QRPA calculations at the CHFB states
by using our 3D FAM-QRPA code developed in Ref.~\cite{washiyama17}.
The full quasiparticle basis states 
that are included in solving the CHFB equation
are used to solve the FAM equation~\eqref{eq:FAM_XY}.
We solve the FAM equation at each $\omega$
up to $\omega=40$\,MeV with a spacing of 0.25\,MeV
and with a smearing width of $\Gamma/2=0.5$\,MeV.

\section{Results and discussion}\label{result}

\begin{figure}
\includegraphics[width=0.7\linewidth,clip]{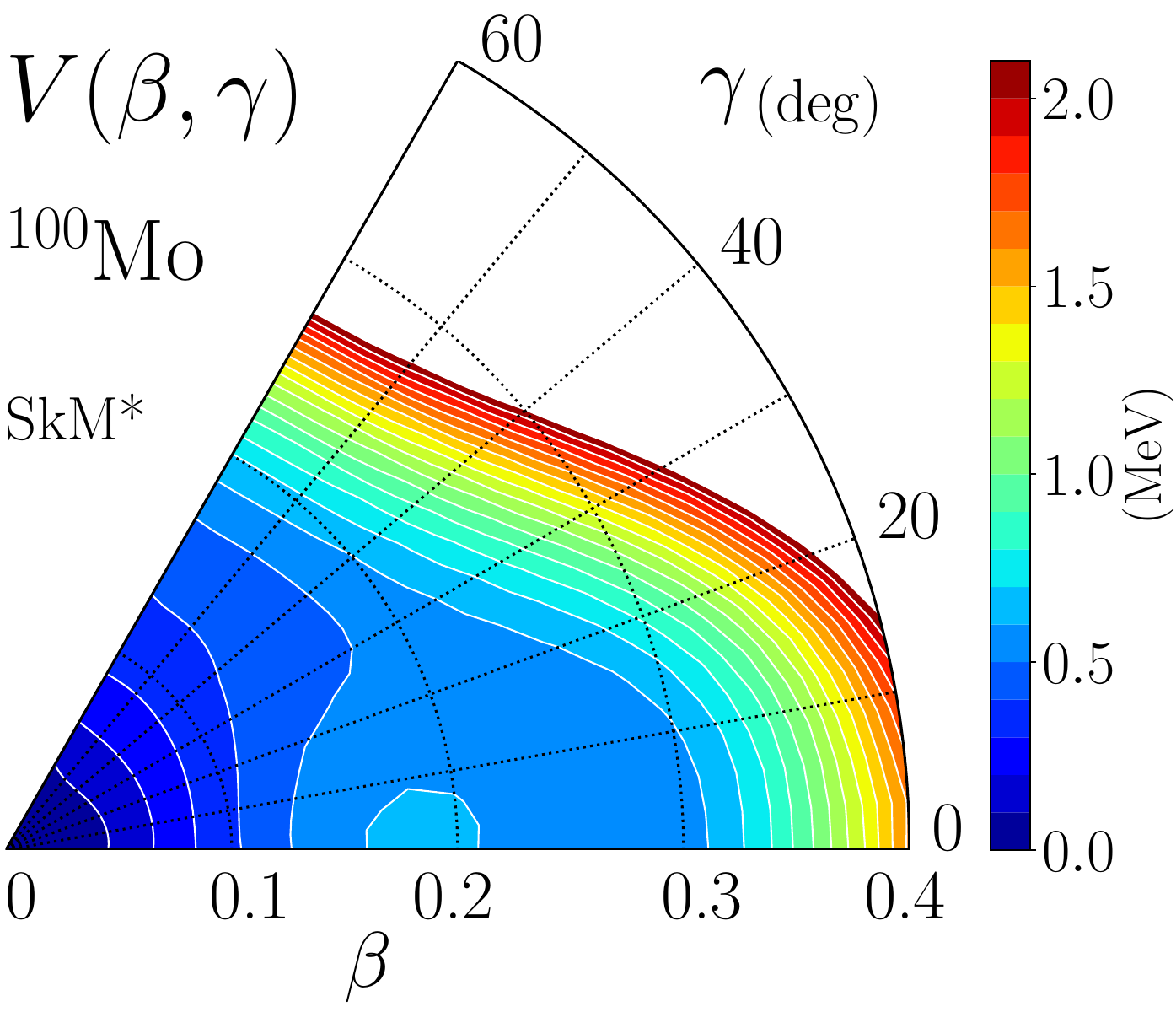}
\caption{
Potential energy surface in the $\beta$--$\gamma$ plane obtained by the CHFB calculation with the SkM$^*$ EDF in $^{100}$Mo.
        }\label{fig:PES}
\end{figure}

Figure~\ref{fig:PES} shows the potential energy surface (PES) in the $\beta$--$\gamma$ plane at $\beta < 0.4$ obtained 
by the CHFB calculation in $^{100}$Mo.
The PES shows the energy minimum at the spherical shape 
and a flat behavior along both the $\beta$ and $\gamma$ directions at $\beta \le 0.3.$
This behavior is consistent with that found in Ref.~\cite{Shi_2023}.

\begin{figure}
\includegraphics[width=\linewidth,clip]{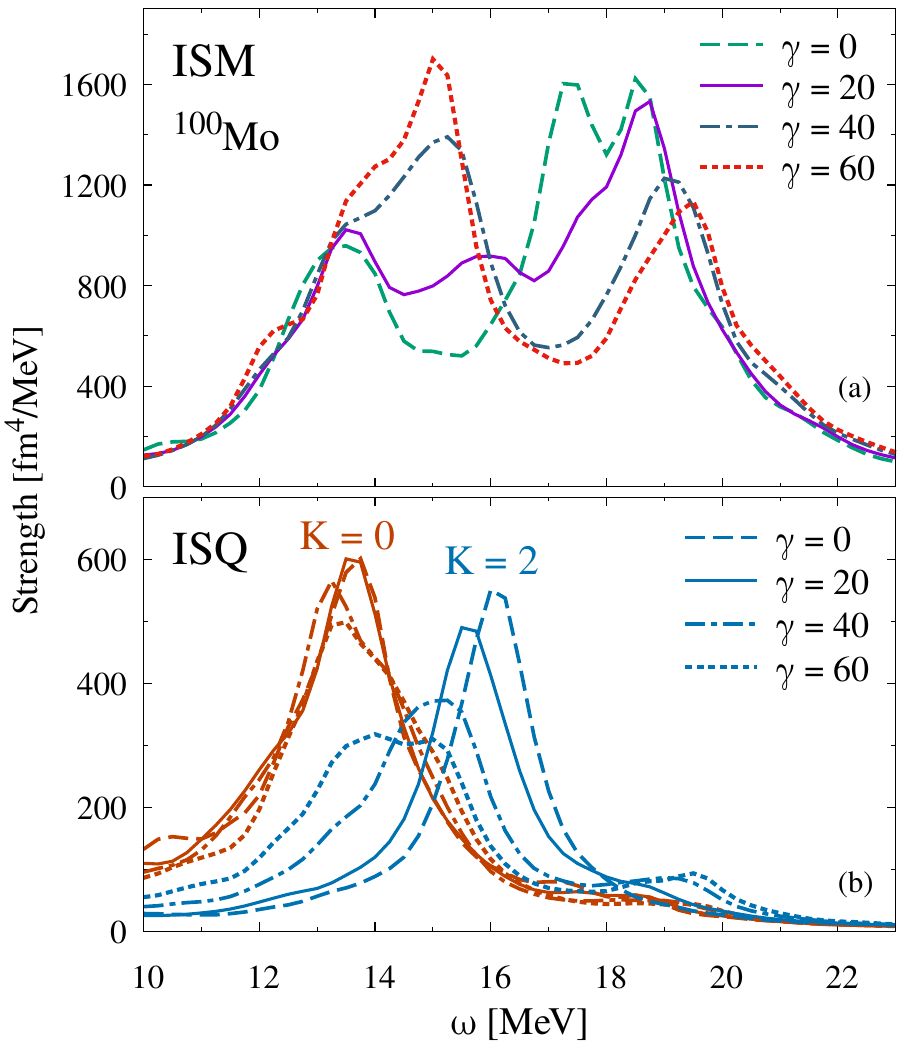}
\caption{
(a) ISM strength and (b) ISQ strength as a function of $\omega$ for different $\gamma$ in $^{100}$Mo with $\beta=0.28$.
}\label{fig:ISM-ISQstrength_gamma0-60}
\end{figure}

Given the $\gamma$-soft property of the potential energy surface (PES) in $^{100}$Mo, 
our next step involves investigating the evolution of 
the ISGMR
as the triaxiality parameter, $\gamma$, increases in $^{100}$Mo. 
To accomplish this, we generate CHFB states at a fixed $\beta=0.28$ with varying $\gamma$ values 
and subsequently conduct FAM-QRPA calculations 
based on these prepared CHFB states.
Figure~\ref{fig:ISM-ISQstrength_gamma0-60}(a) displays 
the ISM strength obtained by systematically altering $\gamma$ in the FAM calculation on the CHFB states with $\beta=0.28$ in $^{100}$Mo. 
With $\gamma=0$, we observe two prominent peaks at $\omega \approx 13$\,MeV and at $\omega \approx 18$\,MeV. 
As the value of $\gamma$ increases, 
an additional peak emerges around $\omega=15$\,MeV, 
positioned between the existing lower and higher peaks. Furthermore, the strength of this middle peak exhibits a noticeable increase.

To understand the evolution of the middle peak with increasing $\gamma$,
Fig.~\ref{fig:ISM-ISQstrength_gamma0-60}(b) presents the $K=0$ and $K=2$ components of the ISQ strength 
obtained by varying $\gamma$.
The peak energy and its shape of the ISQ $K=0$ strength 
do not much depend on $\gamma$,
while the peak energy of the ISQ $K=2$ strength decreases 
and its width becomes broader as $\gamma$ increases.
Notably, the peak energy of the ISQ $K=2$ strength 
aligns with that of the middle peak in the ISM strength.
This finding corroborates the results from a previous study~\cite{Shi_2023}, which demonstrated the appearance of an additional peak or plateau in the ISM strength at $\beta=0.28$, $\gamma=20^\circ$ and $30^\circ$ in $^{100}$Mo using the SkM$^*$ EDF. 
Our results indicate that this coincidence persists 
for $\gamma > 30^\circ$ as well.

\begin{figure}
\includegraphics[width=0.9\linewidth,clip]{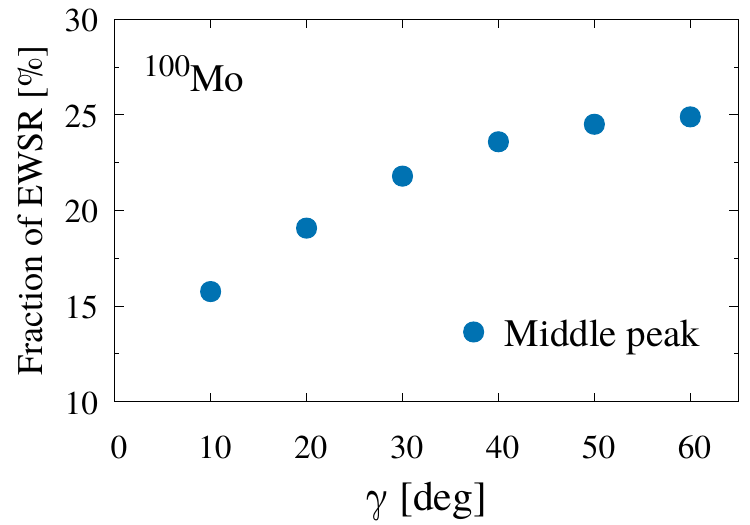}
\caption{
Fraction of the EWSR value for the middle peak in the ISM strength as a function of $\gamma$.
        }\label{fig:EWSR_ISM_middlepeak}
\end{figure}

To quantitatively observe the evolution of the middle peak in the ISM strength, 
we utilize the fraction of the energy-weighted sum rule (EWSR) value for the middle peak. 
We define the energy interval, denoted as $E_1$ and $E_2$, for the middle peak as follows: 
$E_1$ is set to be the average of the mean energies 
of the ISQ $K=0$ and $K=2$ strengths, 
and $E_2$ is the average of the mean energy of the ISQ $K=2$ strength and the peak energy of the higher component in the ISM strength.
Here, we determined the mean energy of the ISQ $K=0$ and $K=2$ using the expression: 
\begin{equation}\label{eq:meanenergy}
    E_{\text{mean}} = \frac{m_1}{m_0},
\end{equation}
where $m_k$ is the $k$th moment of the strength function, defined as 
\begin{equation}\label{eq:EWSR}
    m_k = \int_{\omega_1}^{\omega_2} d\omega \, \omega^k S(\omega), 
\end{equation}
where $\omega_1=10$ MeV and $\omega_2=23$ MeV determine the energy region of the giant resonance in this nucleus.
The peak energy of the higher component in the ISM strength
is determined by a Lorentzian fit.
Figure~\ref{fig:EWSR_ISM_middlepeak} shows the fraction of the EWSR value for the middle peak in the ISM strength as a function of $\gamma$.
It is clearly shown that the fraction of the EWSR value for the middle peak increases as $\gamma$ increases up to $60^\circ$.
A similar behavior was obtained in a triaxial harmonic oscillator potential model in Ref.~\cite{shimizu84},
where the ISGMR strength at the middle peak 
monotonically increases as $\gamma$ increases 
from $0^\circ$ to $60^\circ$.

\begin{figure}
\includegraphics[width=\linewidth,clip]{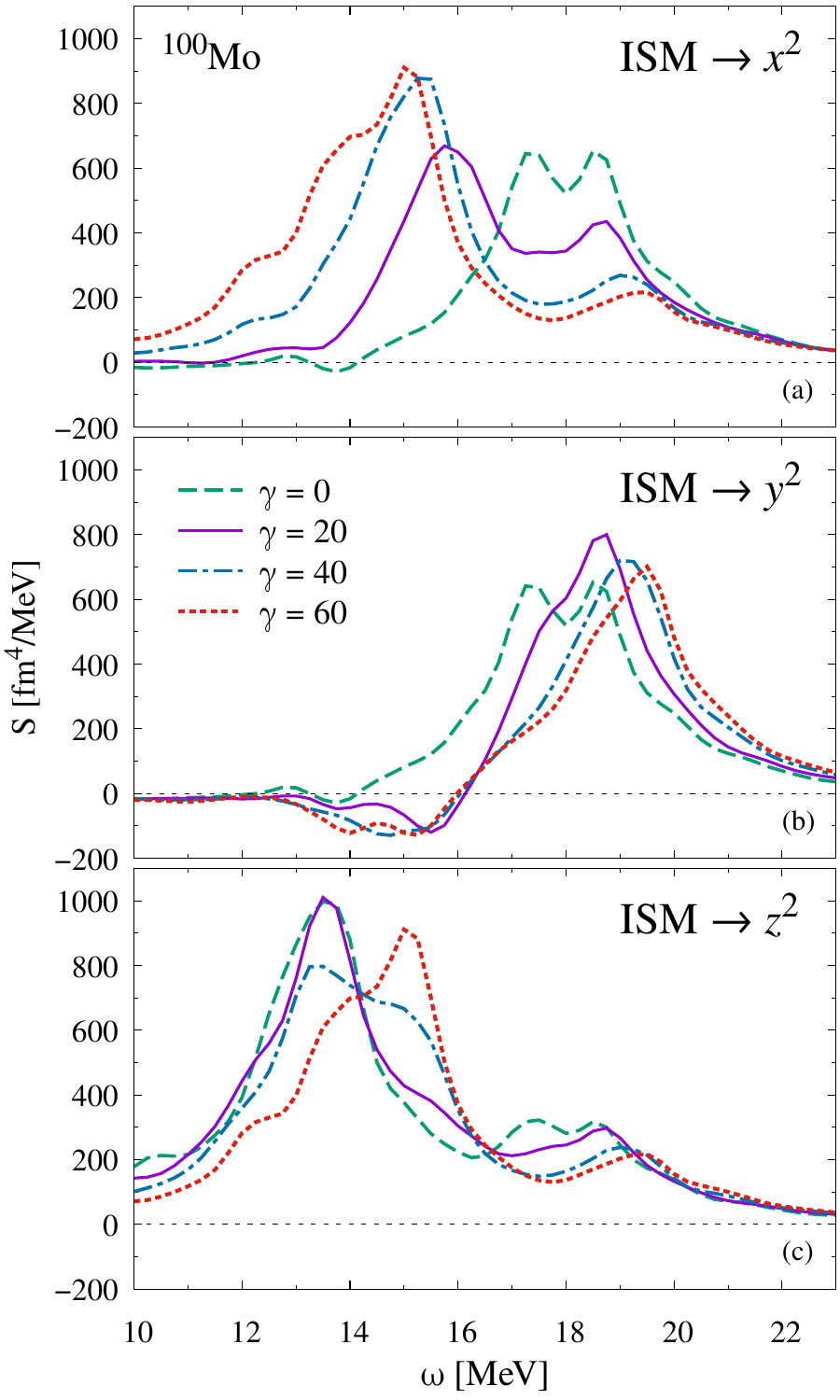}
\caption{
Response functions of ISM $\to x^2$ (a), ISM $\to y^2$ (b) and ISM $\to z^2$ (c) for different $\gamma$ in $^{100}$Mo with $\beta=0.28$.
}\label{fig:strength_x2y2z2_gamma0-60}
\end{figure}

We have obtained the emergence of the middle peak in the ISM strength 
and the agreement between the energy of the middle peak in the ISM strength
and the peak energy of the ISQ $K=2$ strength.
The authors of Ref.~\cite{Shi_2023} concluded that 
the appearance of the middle
peak is attributed to 
a coupling between the ISM and ISQ $K=2$ modes
and  
anticipated that 
at triaxial shapes with nonzero $\braket{Q_{22}^{(+)}}$, namely $\braket{x^2}\neq\braket{y^2}$, in the ground state,
the external ISM field induces different vibrations in the $x$ and $y$ directions.

To see what really happens,
we investigate vibrations of matter distributions in $x$, $y$, and $z$ directions by the external ISM field
that is isotropic.
We calculate the FAM response function of $\hat{F^\prime} =x^2$, $y^2$, or $z^2$ and the ISM field ($\hat{F}=\hat{r}^2$).
Figure~\ref{fig:strength_x2y2z2_gamma0-60} shows 
the imaginary part of 
these response functions,
\begin{equation}\label{eq:Im_response_function}
    S_{\hat{F}^\prime \hat{F}}(\omega) = -\frac{1}{\pi} \text{Im} R_{\hat{F}^\prime\hat{F}}(\omega),
\end{equation}
denoted as ISM $\to x^2$, ISM $\to y^2$, and ISM $\to z^2$, respectively, for different $\gamma$.
Note again that a shape with $\braket{z^2} \ge \braket{x^2} \ge \braket{y^2}$ is obtained at $0^\circ \le \gamma \le 60^\circ$
in our convention.
At $\gamma=0^\circ$,
the response of
ISM $\to x^2$ is identical to that of ISM $\to y^2$ because of the axially symmetric shape.
At $\omega\approx 13.5$\,MeV,
corresponding to the low peak in the ISM strength,
the response of 
the ISM $\to z^2$ is significantly higher than those of ISM $\to x^2$ and ISM $\to y^2$.
When the axial symmetry is broken,
a discernible difference between those of 
ISM$\to x^2$ and ISM $\to y^2$ emerges, 
leading to a finite ISQ $K=2$ strength, which is proportional to $x^2-y^2$,
induced by the ISM field.
Furthermore, the peak energy of ISM $\to x^2$ becomes lower, and its response at $\omega \approx 15$\,MeV, corresponding to the middle peak energy in the ISM strength,
becomes higher
while the response at $\omega\approx 18$\,MeV
decreases
as $\gamma$ increases from $0^\circ$ to $60^\circ$. 
For ISM $\to z^2$,
the response at $\omega\approx 15$\,MeV becomes higher
while the response at $\omega\approx 13.5$\,MeV becomes lower
when $\gamma$ increases from $0^\circ$ to $60^\circ$.
At $\gamma=60^\circ$, 
the response of the ISM $\to x^2$ is identical to that of ISM $\to z^2$
due to the axially symmetric shape with the $y$ being the symmetry axis.

The response function in Eq.~\eqref{eq:Im_response_function}
is related to the Fourier transform of vibrations of matter distributions
in $x$, $y$, and $z$ directions induced by the ISM field. 
At $\gamma=0^\circ$,
since
the response of ISM $\to x^2 $ and that of ISM $\to y^2$ are the same,
the vibration of matter distributions in $x$ direction and that in $y$ direction are the same.
Since the response of ISM $\to z^2$ and 
that of ISM $\to x^2$ and ISM $\to y^2$ are different,
the vibration of matter distributions in $z$ direction and that in $x$ and $y$ directions are different in $\gamma=0^\circ$.
As $\gamma$ increases, 
the vibrations of matter distributions in $x$, $y$, and $z$ directions
become different.
Eventually, at $\gamma=60^\circ$,
the responses of ISM $\to x^2$ and ISM $\to z^2$ become identical,
indicating that the vibration of matter distributions in $x$ and that in $z$ become identical.
The external multipole fields $F_L$ that we have employed 
are rotational-invariant.
The breaking of the spherical symmetry in the ground state density manifests the asymmetry of the 
strengths in $F_L$ with respect to $x, y$, and $z$ directions.

\begin{figure}
\includegraphics[width=\linewidth,clip]{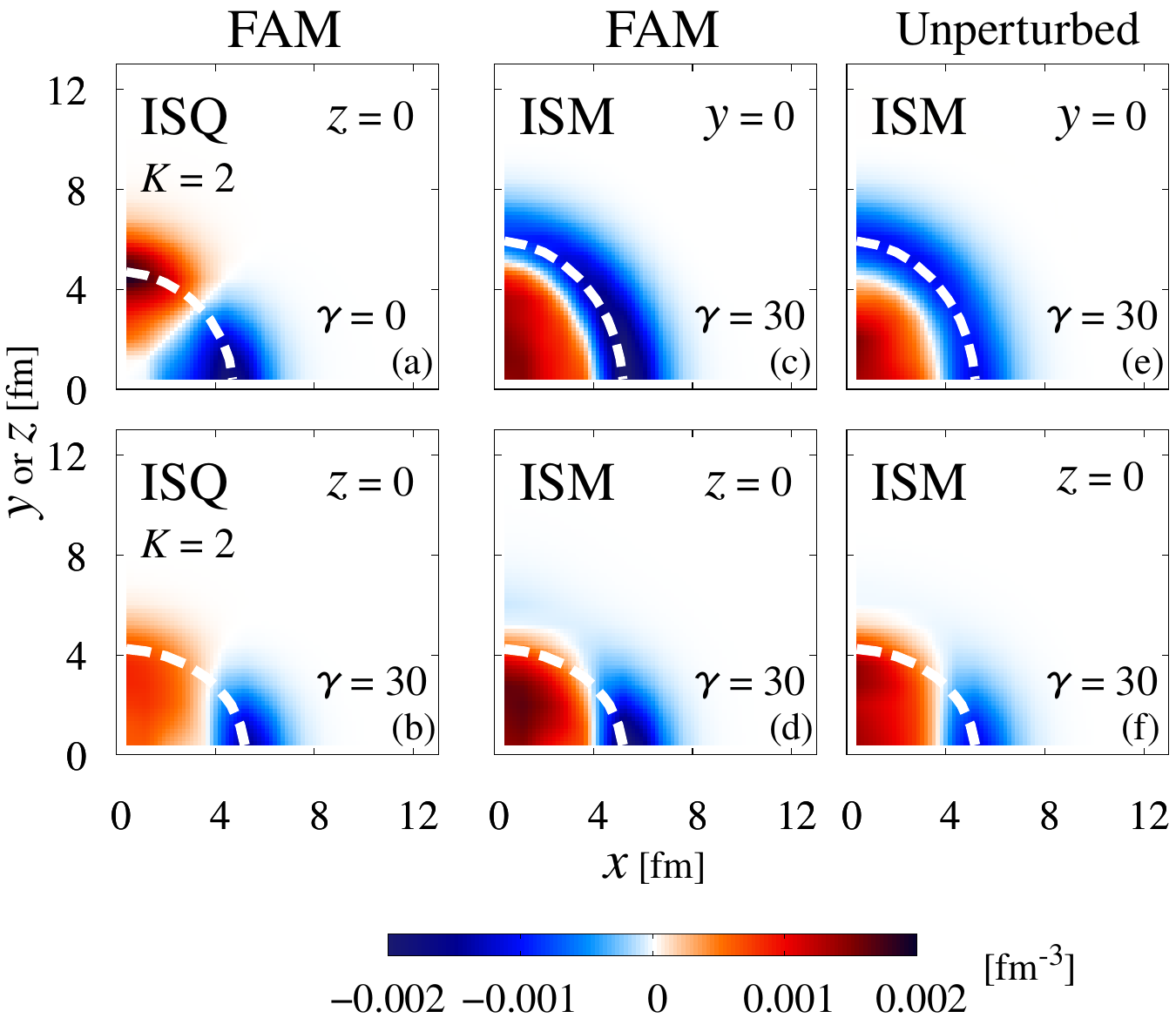}
\caption{
Induced densities for the ISQ $K=2$ field at $\omega=16$\,MeV 
on the $z=0$ plane at
$\gamma=0^\circ$ (a) and $\gamma=30^\circ$ (b),
and for the ISM field at $\omega=16$\,MeV 
on the $y=0$ (c) and $z=0$ plane (d)
and induced densities without the residual ISM field at $\omega=20$\,MeV 
on the $y=0$ (e) and $z=0$ plane (f) 
at $\gamma=30^\circ$ in $^{100}$Mo.
The dashed line indicates the contour of 
the isoscalar density $\rho=0.08$\,fm$^{-3}$ obtained by the CHFB calculation at $\gamma=30^\circ$,
except for (a) at $\gamma=0^\circ$.
}\label{fig:induced_density_ISM_middle_Mo100}
\end{figure}

To understand how the ISM field induces a significant contribution of quadrupole vibrations,
it is useful to analyze the spacial property of the induced density of the collective states.
The induced density at a frequency $\omega$ in the FAM is obtained with the FAM $X$ and $Y$ amplitudes as
\begin{align}\label{eq:FAM_induced_density}
\delta\rho(\bm{r};\omega)&=\sum_{ij} \phi_{i}(\bm{r})[UX(\omega)V^T + V^*Y^T(\omega)U^\dagger]_{ij}\phi_j^*(\bm{r}),
\end{align}
where $U$ and $V$ are the Bogoliubov transformation matrices and $\phi_i(\bm{r})$ are the HF basis wave functions. 
Figure~\ref{fig:induced_density_ISM_middle_Mo100}(c) and (d) show the imaginary part of 
the induced density on the $y=0$ and $z=0$ planes, respectively, by the ISM field at $\omega = 16$\,MeV, 
which approximately corresponds to the energy of the middle peak in the ISM strength,
at $\gamma=30^\circ$, $\beta=0.28$ in $^{100}$Mo.
The dashed line indicates the contour of the isoscalar density $\rho=0.08$\,fm$^{-3}$ of this nucleus obtained by the CHFB calculation.
The induced density on the $y=0$ plane (c) shows that
the density becomes higher inside the nucleus and lower at the nuclear surface, 
and shows almost an isotropic nature.
On the other hand,
the induced density on the $z=0$ plane (d) shows 
an anisotropic nature.
The induced density has a node along the $x$ direction and
becomes higher at smaller $x$
and lower at larger $x$.
This spacial property of the induced density resembles that 
in the $K=2$ of the quadrupole vibration mode,
which is shown in Fig.~\ref{fig:induced_density_ISM_middle_Mo100}(b).
For comparison, the induced density for the ISQ $K=2$ at $\gamma=0^\circ$ is shown in (a),
which has a node along $x=y$.

\begin{figure}
\includegraphics[width=0.9\linewidth,clip]{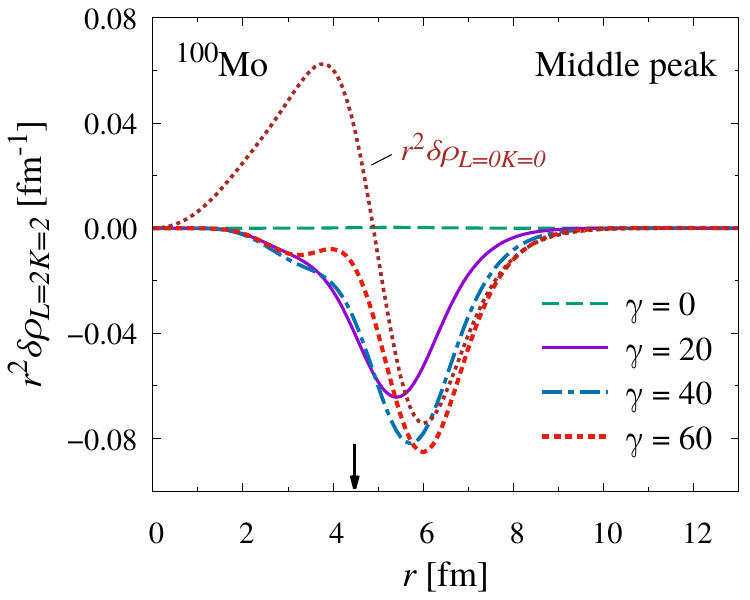} 
\caption{
$L=2$, $K=2$ component of the induced density by the ISM field for different $\gamma$.
The nuclear radius is indicated in the arrow.
The $L=0$ component of the induced density at $\gamma=30^\circ$ is plotted by the dotted line. 
}\label{fig:multipole_expansion_Mo100_r2_gamma0-60}
\end{figure}

We go through the property of the induced density 
at the middle peak.
We show in Fig.~\ref{fig:multipole_expansion_Mo100_r2_gamma0-60} 
the $L=2$, $K=2$ component of the induced density
of the ISM field
for different $\gamma$ obtained by the multipole expansion
of the induced density of the ISM field.
To see the surface property of the induced density,
we plot the induced density multiplied by $r^2$ in the figure.
The $L=2$, $K=2$ component of the induced density vanishes for the axially symmetric case ($\gamma=0^\circ)$.
The magnitude of the $L=2$, $K=2$ component
at the surface becomes larger for larger $\gamma$.
For comparison, we plot the $L=0$ component of the induced density at $\gamma=30^\circ$
by the dotted line. 
Its magnitude at the surface region
is comparable to that in the $L=2$, $K=2$ component at $\gamma=20^\circ$, $40^\circ$, and $60^\circ$.
This clearly shows a strong mixing of the 
$L=2$, $K=2$ component to the ISM strength
at the middle peak energy.

\begin{figure}
\includegraphics[width=\linewidth]{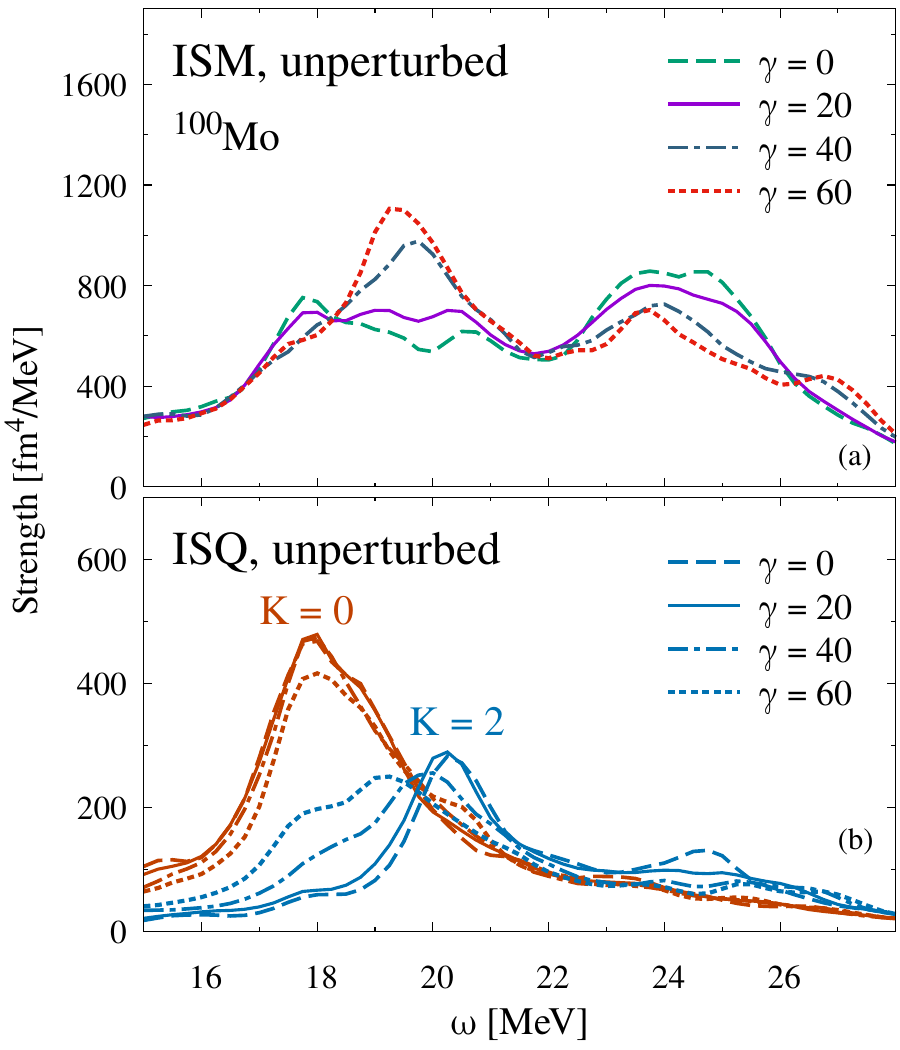}
\caption{
Same as Fig.~\ref{fig:ISM-ISQstrength_gamma0-60},
but for the unperturbed strength.
}
\label{fig:ISM-ISQstrength_unperturbed_gamma0-60}
\end{figure}

To ascertain whether the origin of the $L=2$, $K=2$ component to the ISM strength
stems from either dynamical effect
or static effect,
we also look into the ISM strength without the residual fields in the FAM calculation, namely the unperturbed strength.
Figure~\ref{fig:ISM-ISQstrength_unperturbed_gamma0-60}
shows the unperturbed ISM and ISQ strengths
at different $\gamma$.
The occurrence of the middle peak in the unperturbed ISM strength at finite $\gamma$ is seen at $\omega\approx 20$\,MeV.
Furthermore, the middle peak in the unperturbed ISM strength
and the peak in the unperturbed ISQ $K=2$ strength
coincide in energy.
These properties are preserved when the residual effects 
are taken into account,
which we have seen in Fig.~\ref{fig:ISM-ISQstrength_gamma0-60}.
A similar feature has been pointed out in the case of the 
isovector monopole excitations in axially-deformed nuclei~\cite{Yoshida:2021gla}.
In addition to this, 
as shown in Fig.~\ref{fig:induced_density_ISM_middle_Mo100}(e) and (f),
the spacial property of the induced density without the residual fields
is almost the same as that with the residual fields shown in Fig.~\ref{fig:induced_density_ISM_middle_Mo100}(c) and (d).
From this analysis,
the triaxial deformation in the ground state density 
induces significant contribution of 
$L=2$, $K=2$ component to the ISM strength.

\section{Summary}\label{summary}

We have investigated the mechanism underlying 
the emergence of the middle peak in the ISM strengths 
in deformed nuclei breaking the axial symmetry. 
We performed the CHFB + FAM-QRPA calculations 
to obtain the ISM and ISQ strengths in the $^{100}$Mo nucleus, 
where the triaxial deformation has been pointed out to appear.
The evolution of the middle peak in the ISM strength is clearly shown in the analysis of the fraction of the EWSR value. 
We also investigated the vibrations of matter distributions in the $x$, $y$, and $z$ directions by the ISM field and 
found that 
the origin of the middle peak in the ISM strength is due to different vibrations of 
matter distributions in $x$ and $y$ directions with $\gamma > 0^\circ$.
The induced density of the ISM field 
possesses an enhanced $L=2$, $K=2$ component with $\gamma >0^\circ$ at the middle-peak energy.
Additionally, we have conducted similar calculations 
for the case without the residual ISM field
and found 
the occurrence of the middle peak in the unperturbed ISM strength
as well as the mixing of the $L=2$, $K=2$ component 
to the ISM strength at the middle peak energy. 
The mixing between the monopole and quadrupole modes is 
given by the ground state deformation. Therefore, the ISM strength of the middle peak becomes strong as $\gamma$ increases.

\section*{Acknowledgments}

The authors thank Umesh Garg for the discussion.
This work was supported by the JSPS KAKENHI (Grant No. JP19K03824 and No. JP20K03943). 
Numerical calculations were performed using computational resources of Wisteria/BDEC-01 Odyssey (the University of Tokyo), provided by the Multidisciplinary Cooperative Research Program in the Center for Computational Sciences, University of Tsukuba.

%

\end{document}